\documentclass{ws-ijgmmp}

\usepackage{xcolor}
\usepackage[verbose,hypertexnames=false]{hyperref}
\hypersetup{colorlinks=false,allbordercolors=blue,pdfborderstyle={/S/U/W 1}}
\usepackage{latexsym}
\begin{document}

\markboth{I. A. Rather, G. Panotopoulos }
{4D EGB with hyperons and phase transition}

%
\catchline{}{}{}{}{}
%

\title{Impact of hyperons on structural properties of neutron stars and hybrid stars within the regularized four-dimensional Einstein-Gauss-Bonnet gravity
}

\author{Ishfaq Ahmad Rather}

\address{Institut f\"{u}r Theoretische Physik, Goethe Universit\"{a}t, \\
Max-von-Laue-Str.~1, D-60438 Frankfurt am Main, Germany\\
\email{rather@astro.uni-frankfurt.de} }

\author{Grigoris Panotopoulos\footnote{Corresponding author}}

\address{Departamento de Ciencias F{\'i}sicas, Universidad de la Frontera,\\
Casilla 54-D, 4811186 Temuco, Chile\\
grigorios.panotopoulos@ufrontera.cl }

\maketitle

\begin{history}
\received{(Day Month Year)}
\revised{(Day Month Year)}
\accepted{(Day Month Year)}
\published{(Day Month Year)}
\end{history}

\begin{abstract}
We investigate the impact of hyperons and phase transition to quark matter on the structural properties of neutron stars within the regularized four-dimensional Einstein-Gauss-Bonnet gravity (4DEGB). We employ the density-dependent relativistic mean-field model (DDME2) for the hadronic phase and the density-dependent quark mass (DDQM) model for the quark phase to construct hadronic and hybrid equations-of-state (EoSs) that are consistent with the astrophysical constraints. The presence of hyperons softens the EoS and with a phase transition, the EoS further softens, and the speed of sound squared drops to around 0.2 for the maximum mass configuration, which lies in the pure quark phase. Adjusting the Gauss-Bonnet coupling constant, $\alpha$, within its allowed range results in a decrease in the mass-radius relationship for negative $\alpha$, and an increase for positive $\alpha$. In addition, functions are fitted to the maximum mass and its associated radius as a function of the constant $\alpha$ to observe its impact on these properties. We find that positive values of $\alpha$ support massive stars consistent with the 2\,$M_{\odot}$ constraint and NICER measurements, while negative values, although compatible with low-mass radius observations, fail to reach the observed maximum mass, particularly for EoSs involving phase transitions. Therefore, astrophysical observations may be used to effectively constrain the allowed range of $\alpha$.
\end{abstract}

\keywords{Relativistic stars; Theories of gravity other than GR; Equation-of-State; Phase transition.}

\ccode{Mathematics Subject Classification 2020: 81Q10, 81Q15, 35J10}

\section{Introduction}\label{sec1}

The question ``How many dimensions are there?'' is one of the fundamental questions modern theoretical physics tries to answer. This is due to the fact that although our observable Universe is clearly four-dimensional ($4D$), advances in High Energy Physics from the 20´s as well as over the last decades, such as Kaluza-Klein theories \cite{Kaluza:1921tu,Klein:1926tv}, Supergravity \cite{Nilles:1983ge} and Superstring/M-Theory \cite{SST1,SST2} suggest that extra spatial dimensions might exist. In addition to that, in more than four dimensions higher order curvature terms are natural in Lovelock theory \cite{Lovelock:1971yv}, while higher order curvature corrections appear in the low-energy effective actions of Superstring Theory \cite{Corley:2001hg}.

\smallskip

As is well-known, in four-dimensional space-times the Gauss-Bonnet (GB) term is a topological invariant, and as such it does not contribute to the equations of motion. However, a few years ago a novel 4$D$ Einstein-Gauss-Bonnet (EGB) theory of gravity was proposed by Glavan and Lin \cite{Glavan:2019inb}. They showed that even in four dimensions there are non-trivial effects coming from the presence of the GB term. The authors' basic idea was to rescale the GB coupling constant by $\alpha \to \alpha/(D -4)$ in $D$ dimensions, and then take the limit $D \to 4$. As a consequence of the procedure, it turned out that one can bypass the conclusions of Lovelock's theorem, and thus avoid the Ostrogradsky instability \cite{Woodard:2015zca}. The resultant theory is now dubbed as the novel 4$D$ EGB theory, and it attracted a lot of attention for some time. Since then, a number of black hole solutions have been found within that theory, and their physical properties have been investigated. For an incomplete list see e.g.
\cite{Ghosh:2020syx,Konoplya:2020juj,Singh:2020xju,HosseiniMansoori:2020yfj,Singh:2020nwo,Wei:2020poh,Yang:2020jno}, charged black holes \cite{Fernandes:2020rpa,Zhang:2020sjh}, black holes coupled to magnetic charges and non-linear Electrodynamics \cite{Jusufi:2020qyw,Abdujabbarov:2020jla,Jafarzade:2020ilt}. Besides that, deflection of light by black holes 
\cite{Islam:2020xmy,Jin:2020emq}, quasi-normal modes \cite{Churilova:2020aca,Aragon:2020qdc}
and shadow casts by black holes \cite{Konoplya:2020bxa,Guo:2020zmf,Zeng:2020dco} have been investigated as well. Moreover, Morris-Thorne-like wormholes and thin-shell wormholes have been investigated in  \cite{Jusufi:2020yus,Liu:2020yhu}.  
In Ref. \cite{Cao:2021nng} it was shown that in order to have a well-defined linearized theory, the geometry must be (locally) conformally flat. In view of the importance of that theory, research has been conducted in order to constrain the new re-scaled Gauss-Bonnet parameter using observational data, see for instance \cite{Wang:2021kuw} and references therein. 

\smallskip

Despite the initial wave of publications on the topic, some criticism, as well as a number of objections, to this approach were subsequently raised \cite{Critica1, Critica2, Critica3, Gurses:2020rxb,Arrechea:2020gjw}. The shortcomings of the original "novel" $4D$ EGB gravity were quickly addressed when it was shown that a well-defined limit $D \rightarrow 4$ can be taken in the gravitational action \cite{Fernandes, Mann1}. In the first work the method consists of introducing a counterterm into the action that is sufficient to cancel the divergence that would otherwise occur. On the contrary, the authors of the second work employed compactification of higher-dimensional spaces a la Kaluza-Klein, generalizing an earlier procedure employed in obtaining the $D \rightarrow 2$ limit of GR \cite{Ross}. Since the derivation of a well-defined limit for $4D$ Einstein-Gauss-Bonnet gravity coupled to a scalar field, there has been interest in testing it as an alternative to Einstein’s General Relativity \cite{Kumar:2020sag, Mishra:2020gce, Clifton:2020xhc, Feng:2020duo}. Using the Tolman-Oppenheimer-Volkoff (TOV) equations modified for the regularized 4DEGB gravity, the authors model stellar structure of quark stars and white dwarf stars in recent works \cite{Mann2, Mann3, Ayan}.

Neutron stars, with radii around 10 km and masses of 2 solar masses or more, are composed of extremely dense hadronic matter, making them ideal for examining cold, dense nuclear matter. Theoretical predictions of their composition and properties like mass and radius depend on the nuclear equation of state (EoS). Neutron stars contain protons to maintain chemical equilibrium, but due to the strong interaction's nonperturbative nature, the exact EoS is not well understood. This makes the precise structure of neutron stars at high densities, where exotic forms of matter may exist, unknown. Comparing theoretical models to astrophysical data helps to clarify the real EoS of cold dense nuclear matter. Theoretical models of neutron stars usually include not just nucleons but also hyperons to account for energy considerations \cite{Glendenning:1997wn}. However, adding hyperons tends to soften the EoS, reducing the neutron star's maximum possible mass and contributing to the ``hyperon puzzle''\cite{2017hspp.confj1002B}.

\smallskip

At extremely high densities, a transition from hadronic matter to quark matter could happen, leading to a hybrid star with a quark core and hadronic shell. This hadron-quark transition is a significant prediction of QCD under extreme conditions. Gravitational waves from neutron star mergers provide important information to help understand the EoS and provide insights into neutron star interiors, including possible quark matter. The oscillations of neutron stars, observable in gravitational wave frequencies, reveal details about the internal structure and possible presence of quark matter, aiding our understanding of dense cosmic matter \cite{PhysRevLett.119.161101, Capano2020}.

\smallskip

The stellar properties for the NS obtained from the EoSs for several different models for the hadronic matter as well as quark matter need to satisfy several astrophysical constraints for the mass and radii. The most important constant on the EoS comes from the mass measurements of massive pulsars such as J1614-2230 ($M=1.97\pm 0.04~M_\odot$) \cite{Demorest:2010bx}, PSR J0348+0432 ($M=2.01\pm 0.04~M_\odot$) \cite{Antoniadis:2013pzd}, and PSR J0740+6620 ($M=2.08\pm 0.07~M_\odot$) \cite{Fonseca:2021wxt} which suggest that the EoS of dense matter must be able to describe neutron stars of $2\,M_{\odot}$ implying that the matter should remain sufficiently stiff at high densities. Additionally, X-ray measurements of the radius of pulsars from the Neutron Star Interior Composition Explorer (NICER) and X-ray Multi-Mirror (XMM) constrain the radius to
$R = 12.39^{+1.30}_{-0.98}$ km \cite{Riley:2021pdl} and $R = 13.7^{+2.6}_{-1.5}$ km \cite{miller2021}. Recent analysis provided an updated values of these measurements to $R = 12.49^{+1.28}_{-0.88}$ \cite{Salmi:2024aum}, and  $R = 12.76^{+1.49}_{-1.02}$ \cite{dittmann2024precisemeasurementradiuspsr}.  The radius measurements for the PSR J0030+0451 at 1.4\,$M_{\odot}$ are, $R = 13.02^{+1.24}_{-1.06}$\,km by \cite{Miller_2019a} and $R = 12.71^{+1.14}_{-1.19}$\,km by \cite{Riley_2019}. These two limits include the recent measurement by \cite{Vinciguerra:2023qxq}. The mass-radius measurement of PSR J0437-4715 to $R = 11.36^{+0.95}_{-0.63}$ km for a mass of $M = 1.418 \pm 0.037~M_{\odot}$ by \cite{Choudhury:2024xbk} favors softer EoS models. Moreover, low mass and radius measurements of the object such as HESS J1731-347 with very low mass $M = 0.77^{+0.20}_{-0.17}~M_{\odot}$ and radius $R = 10.4^{+0.86}_{-0.78}$ km \cite{Doroshenko2022}, and XTE J1814-338 with a mass and radius of $M = 1.21^{+0.05}_{-0.05}~M_{\odot}$ and $R = 7.^{+0.4}_{-0.4}$ km \cite{Kini:2024ggu} challenge existing EoS models. The EoS not fulfilling such constraints, especially the 2\,$M_{\odot}$ are ruled out.

\smallskip

In the present work, we propose to investigate some astrophysical implications of the regularized $4D$ Einstein-Gauss-Bonnet gravity. To be more precise, we shall study the impact of the Gauss-Bonnet coupling constant $\alpha$ on the mass-radius relationships of hadronic and hybrid stars, including hyperons. In this initial investigation, we restrict our analysis to the simple case of static, spherically symmetric, and isotropic stellar configurations, with an extension to pressure anisotropy. This allows for the use of the modified Tolman-Oppenheimer-Volkoff (TOV) formalism, providing a clear baseline for understanding the fundamental impact of the Gauss-Bonnet coupling on stellar structure. We acknowledge that realistic astrophysical neutron stars rotate, and they may exhibit magnetic anisotropies in their cores. The inclusion of those effects requires a significant extension of the theoretical framework beyond the modified TOV equations to a system of coupled partial differential equations, a formidable challenge within 4DEGB gravity that we postpone for future work.

The plan of our work is as follows: In Section \ref{sec2}, we present the theory and the structure equations for non-rotating stars. In Section \ref{sec:eos}, we discuss the equation of state (EoS) of hadronic matter, deconfined quark matter, and phase transition. Section \ref{results} presents and discusses our main numerical results for different EoS with different GB constant values. We also discuss the effect of the pressure anisotropy parameter on the maximum mass. Finally, we summarize our work in section \ref{summary}. Throughout the manuscript, we adopt the mostly positive metric signature, and we work in geometrical units where Newton's constant and the speed of light in vacuum are set to unity, $G=1=c$.

 \section{Relativistic Stars within the regularized 4$D$ EGB Gravity} \label{sec2}

\subsection{Action and field equations}

In the present work we are interested in astrophysical implications of what is now called the regularized Einstein-Gauss-Bonnet in four dimensions \cite{Fernandes, Mann1}. The gravitational part of the action is given by the usual Einstein-Hilbert term plus the Gauss-Bonnet term coupled to a scalar field $\phi$
\begin{equation}
S = S_{EH} + S_{4}^{GB} + S_{\text{M}}[g]
\end{equation}
\begin{equation}
S_{EH} = \int  d^{4} x \sqrt{-g} \frac{R}{\kappa^2} 
\end{equation}
\begin{equation}
S_{4}^{GB}
=\alpha \int d^{4} x \sqrt{-g}\left[  \phi \mathcal{G}+4 G_{\mu \nu} \nabla^\mu \phi \nabla^\nu \phi-4(\nabla \phi)^2 \square \phi+2(\nabla \phi)^4\right]
\end{equation}
where $S_{\text{M}}[g]$ is the action of the matter content, $g$ is the determinant of the metric tensor $g_{\mu\nu}$, $\kappa$ is the gravitational constant, the Gauss-Bonnet coupling constant, $\alpha$, has dimensions of $[length]^2$, while $\mathcal{G}$ is the Guass-Bonnet term defined by
\begin{equation}
\mathcal{G} \equiv R^{\mu\nu\rho\sigma} R_{\mu\nu\rho\sigma}- 4 R^{\mu\nu}R_{\mu\nu}+ R^2\label{GB}.
\end{equation}
where $R$ is the Ricci scalar, while $R_{\mu\nu}$ and $R_{\mu\sigma\nu\rho}$ are the Ricci tensor and the Riemann tensor, respectively.  This action belongs to the Horndeski class of theories \cite{Horndeski:1974wa, Lu:2020iav, Kobayashi:2020wqy}, where $G_2=8 \alpha X^2$, $G_3=8 \alpha X$, $G_4=1+4 \alpha X$ and $G_5=4 \alpha ln(X)$ \cite{Fernandes} with $X=-(1/2) (\partial \phi)^2$ being the standard kinetic term of the scalar field.

\smallskip

The field equations are obtained from a straightforward variational principle applied to the total action of the system. Variation with respect to the scalar field yields

\begin{equation}\label{eq:eomscalar}
\begin{aligned}
0 =&-\mathcal{G}+8 G^{\mu \nu} \nabla_{\nu} \nabla_{\mu} \phi+8 R^{\mu \nu} \nabla_{\mu} \phi \nabla_{\nu} \phi-8(\square \phi)^{2} \\
&+8(\nabla \phi)^{2} \square \phi+16 \nabla^{a} \phi \nabla^{\nu} \phi \nabla_{\nu} \nabla_{\mu} \phi  +8 \nabla_{\nu} \nabla_{\mu} \phi \nabla^{\nu} \nabla^{\mu} \phi 
\end{aligned}
\end{equation}
while the variation with respect to the metric tensor yields

\begin{align}\label{eq:eommetric}
\kappa^2 T_{\mu \nu} &=  G_{\mu \nu} + \alpha \bigg[ 
\phi H_{\mu \nu} - 2 R \left( \nabla_\mu \phi \nabla_\nu \phi + \nabla_\nu \nabla_\mu \phi \right) \\
&+ 8 R_{(\mu}^{\sigma} \nabla_{\nu)} \nabla_\sigma \phi 
+ 8 R_{(\mu}^{\sigma} \nabla_{\nu)} \phi \nabla_\sigma \phi \notag \\
& - 2 G_{\mu \nu} \left( (\nabla \phi)^2 + 2 \square \phi \right)
- 4 \left( \nabla_\mu \phi \nabla_\nu \phi + \nabla_\nu \nabla_\mu \phi \right) \square \phi \notag \\
& - \left[ g_{\mu \nu} (\nabla \phi)^2 - 4 \nabla_\mu \phi \nabla_\nu \phi \right] (\nabla \phi)^2 
+ 8 \nabla_{(\mu} \phi \nabla_{\nu)} \nabla_\sigma \phi \nabla^\sigma \phi \notag \\
& - 4 g_{\mu \nu} R^{\sigma \rho} \left( \nabla_\sigma \nabla_\rho \phi + \nabla_\sigma \phi \nabla_\rho \phi \right) 
+ 2 g_{\mu \nu} (\square \phi)^2 \notag \\
& - 4 g_{\mu \nu} \nabla^\sigma \phi \nabla^\rho \phi \nabla_\sigma \nabla_\rho \phi 
+ 4 \nabla_\sigma \nabla_\nu \phi \nabla^\sigma \nabla_\mu \phi \notag \\
& - 2 g_{\mu \nu} \nabla_\sigma \nabla_\rho \phi \nabla^\sigma \nabla^\rho \phi 
+ 4 R_{\mu \nu \sigma \rho} \left( \nabla^\sigma \phi \nabla^\rho \phi + \nabla^\rho \nabla^\sigma \phi \right) 
\bigg]
\end{align}

where  the Einstein tensor, $G_{\mu\nu}$, and Lanczos tensor,
$H_{\mu\nu}$, are defined by
\begin{eqnarray}
&& G_{\mu\nu} \equiv R_{\mu\nu}-\frac{1}{2}R~ g_{\mu\nu},\nonumber\\
&& H_{\mu\nu} \equiv 2\Bigr( R R_{\mu\nu}-2R_{\mu\sigma} {R}{^\sigma}_{\nu}  \nonumber \\
&& -2 R_{\mu\sigma\nu\rho}{R}^{\sigma\rho} - R_{\mu\sigma\rho\delta}{R}^{\sigma\rho\delta}{_\nu}\Bigl) - \frac{1}{2}~g_{\mu\nu}~\mathcal{G},\label{FieldEq}\\
\end{eqnarray}
while the energy-momentum tensor of matter distribution is defined by
\begin{equation}
T_{\mu\nu} = -\frac{2}{\sqrt{-g}}\frac{\delta\left(\sqrt{-g}\mathcal{L}_M\right)}{\delta g^{\mu\nu}}.
\end{equation}

It is worth mentioning that upon taking the trace, the field equations imply the following relationship
\begin{equation}\label{eq:fieldeqntrace}
\kappa^2 T = -R-\frac{\alpha}{2} \mathcal{G},
\end{equation}
with $T = T_\mu^\mu = g^{\mu \nu}T_{\mu \nu}$ being the trace of the stress-energy tensor of the matter content. The trace equation is exactly of the same form as the trace of the field equations of the original "novel" 4DEGB theory \cite{Glavan:2019inb, Fernandes}. In the limit $\alpha \rightarrow 0$ the trace equation of GR is recovered. Furthermore, in vacuum where $T=0$ it takes the form $-R-\frac{\alpha}{2} \mathcal{G}=0$, and therefore the exterior vacuum solution in Schwarzschild-like coordinates $t, r, \theta, \varphi$ is found to be
\begin{eqnarray}\label{metricExt}
ds^2= - F(r)dt^2 + \frac{1}{F(r)}dr^2 + r^{2}d\Omega_{2}^2,
\end{eqnarray} 
\begin{equation}
F(r) = 1+\frac{r^2}{2 \alpha}\left(1-\sqrt{1+\frac{8 \alpha M}{r^3}}\right)
\end{equation}
where $d\Omega_{2}^2=d \theta^2 + sin^2 \theta \: d \varphi^2$ is the metric of the unit 2-dimensional sphere, and $M$ is the stellar mass. This solution generalizes the Schwarzschild geometry of GR in the absence of matter \cite{SBH}. Clearly, in the limit $\alpha \to 0$ we recover the Schwarzschild solution.
It is clear from \cite{Lu:2020iav, Kobayashi:2020wqy} that the black hole solutions obtained in \cite{Glavan:2019inb} are also solutions to the field equations of the regularized 4DEGB gravity, and therefore not surprisingly the exterior vacuum solution is the same both in the "novel" and in the regularized 4DEGB gravity.


\subsection{Relativistic fluid spheres within regularized 4D EGB}

To obtain spherically symmetric interior solutions describing hydrostatic equilibrium of relativistic stars we choose the metric of the following form in Schwarzschild-like coordinates $t, r, \theta, \varphi$
\begin{eqnarray}\label{metric}
ds^2= - e^{2\Phi(r)}dt^2 + e^{2\Lambda(r)}dr^2 + r^{2}d\Omega_{2}^2,
\end{eqnarray} 
where the metric functions $\Phi(r)$ and $\Lambda(r)$ depend only on the radial coordinate $r$. Here we are interested in non-rotating stars made of isotropic matter, the energy-momentum tensor of which in $(1+3)$-dimensional space-times may be written down as follows 
\begin{equation}\label{emt}
T_{\mu\nu}=(\epsilon+P) u_{\mu} u_{\nu} + P g_{\mu\nu},
\end{equation}
where $u^{\mu}$ is the four-velocity, $\epsilon = \epsilon(r)$ is the energy density, and $P = P(r)$ is the pressure of the fluid.  

Regarding the scalar field, the solution
\begin{equation}
\phi(r)=\int \frac{1 - e^{\Lambda}}{r} dr
\end{equation}
ensures  that \eqref{eq:eomscalar} is automatically satisfied \cite{Mann2, Mann3}. Next, the field equations for the metric tensor take the form 
\begin{equation}\label{DRE1}
\begin{aligned}
\frac{2}{r} \frac{d\Lambda}{dr} &= e^{2\Lambda} 
~ \biggl[8\pi \: \epsilon(r) 
- \frac{1-e^{-2\Lambda}}{r^2}\left(1 - \frac{\alpha(1-e^{-2\Lambda})}{r^2}\right)\biggr] \\
&\quad \times \biggl[1 + \frac{2\alpha(1-e^{-2\Lambda})}{r^2}\biggr]^{-1}, \\ 
\frac{2}{r} \frac{d\Phi}{dr} &= e^{2\Lambda} 
~ \biggl[8\pi \: P(r) 
+ \frac{1-e^{-2\Lambda}}{r^2} \left(1 - \frac{\alpha(1-e^{-2\Lambda})}{r^2}\right)\biggr] \\
&\quad \times \biggl[1 + \frac{2\alpha(1-e^{-2\Lambda})}{r^2}\biggr]^{-1}, \\ 
\frac{dP}{dr} &= - (\epsilon + P) \frac{d\Phi}{dr}.
\end{aligned} 
\end{equation}

If, in the usual way, we define the mass function, $m(r)$, within a sphere of radius $r$ through the relation 
\begin{equation}
\label{Lamdef}
    e^{-2 \Lambda}=1+\frac{r^2}{2 \alpha}\left(1-\sqrt{1+\frac{8 \alpha m(r)}{r^3}}\right)
\end{equation}
we arrive at the 4DEGB modified TOV equations, namely \cite{Mann2, Mann3, Ayan}
\begin{align}
& \frac{d P}{d r}=\frac{(P+\epsilon)\left[r^3(\Gamma+8 \pi \alpha P-1)-2 \alpha m\right]}{r^2 \Gamma\left[r^2(\Gamma-1)-2 \alpha\right]} \label{dpdr}\\
&\frac{d m}{d r}=4 \pi r^2 \epsilon
\label{dmdp}
\end{align}
where we have introduced the function $\Gamma=\sqrt{1+\frac{8 \alpha m}{r^3}}$.

It is not difficult to see that one may recover the standard TOV equations of Einstein's General Relativity for isotropic fluid distributions when $\alpha \to 0$. 

\smallskip

The structure equations can be integrated numerically, imposing the conditions $P(0) = P_c, m(0)=0$ at the center of the fluid sphere, with $P_c$ being the central pressure. To obtain interior solutions describing hydrostatic equilibrium, we need to derive an equation of state, and this is the subject of the next section. We impose the following matching conditions at the surface of the star, $r \to R$
\begin{equation}
P(R) = 0, \; \; \; \; \; m(R)=M,
\end{equation}
with $R$ being the stellar radius. 

 We vary the GB parameter in the range [-5 km$^2$, +5 km$^2$] following \cite{Banerjee:2020dad}. This specific range is chosen not because it reflects the tightest observational constraints, but for illustrative purposes to clearly demonstrate the qualitative impact of both positive and negative coupling constants on the stellar structure. A detailed discussion on how this range relates to current astrophysical constraints is provided in Section 4 after presenting the mass-radius results.

\section{Equation-of-State Formalism}
\label{sec:eos}

\subsection{Hadronic matter}

To analyze how the Gauss-Bonnet (GB) constant $\alpha$ impacts the mass-radius relationship of neutron stars (NS), we utilize a density-dependent relativistic mean-field (DD-RMF) model to describe hadronic matter. This model is acclaimed for accurately replicating the experimental features of nuclear matter and aligning with astrophysical constraints. It involves interactions between nucleons and other hadrons through virtual meson exchanges. The particular DD-RMF model applied includes interactions with the scalar meson $\sigma$, vector mesons $\omega$ and $\phi$, and the isovector-vector meson $\vec{\rho}$.

\smallskip

In any RMF theory, the Lagrangian density serves as the fundamental assumption, incorporating components from free baryons, mesons, and their interaction terms. Through the mean-field approximation, the Lagrangian used for the relativistic model in describing hadronic interactions is represented by
%
\begin{align}
     \mathcal{L}_{\rm DD-RMF}=  {}& 
     \sum_{b\in H}  \bar \psi_b \Big[  i \gamma^\mu\partial_\mu - \gamma^0  \big(g_{\omega b} \omega_0  +  g_{\phi b} \phi_0+ g_{\rho b} I_{3b} \rho_{03}  \big) \nonumber\\
&
- \left( m_b- g_{\sigma b} \sigma_0 \right)  \Big] \psi_b
+\sum_l\Bar{\psi}_l\left(i\gamma^\mu\partial_\mu-m_l\right)\psi_l
\nonumber\\
&
- \frac{1}{2} m_\sigma^2 \sigma_0^2  +\frac{1}{2} m_\omega^2 \omega_0^2 +\frac{1}{2} m_\phi^2 \phi_0^2 +\frac{1}{2} m_\rho^2 \rho_{03}^2 . \label{lagrangian}
\end{align}
%
The first sum in the above equation represents the Dirac-type interacting Lagrangian for the spin-1/2 baryon octet, $H=\{n,p,\Lambda,\Sigma^-,\Sigma^0,\Sigma^+,\Xi^-,\Xi^0\}$. The second term describes the contribution from leptons in the hadronic matter as a free non-interacting fermion gas, $l=\{e,\mu\}$, as their contribution is necessary to ensure the $\beta$-equilibrium and charge neutrality essential to stellar matter. The remaining terms account for the purely mesonic part of the Lagrangian.

In DD-RMF models, coupling constants depend on scalar or vector densities, with vector density commonly parameterized to impact self-energy alone \cite{PhysRevLett.68.3408}. We utilize the DD-RMF parametrization DDME2, where meson couplings scale with the baryonic density factor $\eta = n_B/n_0$, obeying the function
\begin{equation}
    g_{i b} (n_B) = g_{ib} (n_0) \frac{a_i +b_i (\eta + d_i)^2}{a_i +c_i (\eta + d_i)^2} 
\end{equation}
for $i=\sigma, \omega, \phi$ and 
\begin{equation}
    g_{\rho b} (n_B) = g_{ib} (n_0) \exp\left[ - a_\rho \big( \eta -1 \big) \right],
\end{equation}
for $i=\rho$. 

\begin{table}[!ht]
\centering
\caption{The coupling parameters of the DDME2 set.\label{T_1} }
\begin{tabular}{p{0.5cm}p{1.2cm}p{0.8cm}p{0.8cm}p{0.8cm}p{0.8cm}p{0.8cm}}
\hline
$i$ & $m_i(\text{MeV})$ & $a_i$ & $b_i$ & $c_i$ & $d_i$ & $g_{i N} (n_0)$\\
 \hline
 $\sigma$ & 550.1238 & 1.3881 & 1.0943 & 1.7057 & 0.4421 & 10.5396 \\  
 $\omega$ & 783 & 1.3892 & 0.9240 & 1.4620 & 0.4775 & 13.0189  \\
 $\rho$ & 763 & 0.5647 & --- & --- & --- & 7.3672 \\
 \hline
\end{tabular}
\end{table}
\begin{table}[]
\centering
\caption{Predictions to the nuclear matter properties at saturation density for the above DDME2 parameter set.\label{T11} }
\begin{tabular}{c|cc}
\hline 
Quantity & Constraints \cite{dutra2014, Oertel:2016bki} & DDME2\\\hline
$n_0$ ($fm^{-3}$) & 0.148--0.170 & 0.152 \\
 $-B/A$ (MeV) & 15.8--16.5  & 16.4  \\ 
$K_0$ (MeV)& 220--260   &  252  \\
 $S_0$ (MeV) & 31.2--35.0 &  32.3  \\
$L_0$ (MeV) & 38--67 & 51\\
\hline
\end{tabular}
\label{T1}
\end{table}

The coupling parameters for the DDME2 set are shown in Table~\ref{T_1}. The model's parameters are determined using experimental constraints on nuclear matter near its saturation density $n_0$, focusing on the binding energy $B/A$, compressibility modulus $K_0$, symmetry energy $S_0$, and its slope $L_0$, as presented in Table~\ref{T11}. These parameters are tailored for pure nucleonic matter, comprised solely of protons and neutrons. To establish the meson couplings for other hadronic species, we define the ratio $\chi_{ib}=g_{i b}/g_{i N}$ for the baryon coupling relative to the nucleon coupling, with $i = \{\sigma,\omega,\phi,\rho\}$. In this study, hyperons are included in the nucleonic matter, and we follow the methodology of \cite{Lopes1} to calculate their respective $\chi_{ib}$ ratios. This formalism maintains a consistent framework based on symmetry principles, particularly ensuring that the Yukawa coupling terms in the Lagrangian density of DD-RMF models are invariant under SU(3) and SU(6) group transformations. Consequently, the couplings are adjusted to achieve the potentials $U_\Lambda =-28$~MeV, $U_\Sigma= 30$~MeV, and $U_\Xi=-4$~MeV using a single free parameter $\alpha_V$. We choose $\alpha_V=1.0$ for the baryon-meson coupling scheme, which corresponds to an unbroken SU(6) symmetry, and the values of $\chi_{ib}$ are given in Table~\ref{T2} \cite{Lopes1}, considering the isospin projections in the Lagrangian terms \cite{issifu}.

\begin{table}[!ht]
\centering
\caption{Baryon-meson coupling constants  $\chi_{ib}$. } 
\label{T2} 
\begin{tabular}{ c c c c c } 
\hline
 b & $\chi_{\omega b}$ & $\chi_{\sigma b}$ & $I_{3b}\chi_{\rho b}$ & $\chi_{\phi b}$  \\
 \hline
 $\Lambda$ & 2/3 & 0.611 & 0 & 0.471  \\  
  $\Sigma^{-}$,$\Sigma^0$, $\Sigma^{+}$ & 2/3 & 0.467 & $-1$, 0, 1 & -0.471 \\
$\Xi^-$, $\Xi^0$  & 1/3 & 0.284 & $-1/2$, 1/2 & -0.314 \\
  \hline
\end{tabular}
\end{table}

From Eq.~\eqref{lagrangian}, thermodynamic quantities can be calculated in the standard way for RMF models. The baryonic density of a baryon of the species $b$ is given by
\begin{equation}
n_b = \frac{\lambda_b}{2\pi ^{2}}\int_{0}^{{k_F}_b}dk\, k^{2}=\frac{\lambda_b}{6\pi ^{2}}{k_F}_b^{3}, \label{eq:rhobarion}
\end{equation} 
where ${k_F}$ denotes the Fermi momentum since we assume the stellar matter to be at zero temperature and $\lambda_b$ is the spin degeneracy factor.
The effective masses are given by
\begin{equation}
    m_b^\ast =m_b- g_{\sigma b} \sigma_0 .
\end{equation}

\smallskip

Solving the energy-momentum tensor, we obtain the energy density as

\begin{align}\label{1a}
\varepsilon_B={}& \sum_b \frac{\gamma_b}{2\pi^2}\int_0^{{k_{F}}_b} dk k^2 \sqrt{k^2 + {m_b^\ast}^2}
\nonumber\\
&
+ \sum_l \frac{1}{\pi^2}\int_0^{{k_{F}}_l} dk k^2 \sqrt{k^2 + m_l^{2}}+ \frac{m_\sigma^2}{2} \sigma_0^2+\frac{m_\omega^2}{2} \omega_0^2
\nonumber\\
&
+\frac{m_\phi^2}{2} \phi_0^2  + \frac{m_\rho^2}{2} \rho_{03}^2 .
\end{align}

The effective chemical potentials become
\begin{equation}
      \mu_b^\ast = \mu_b- g_{\omega b} \omega_0 - g_{\rho b} I_{3b} \rho_{03} - g_{\phi b} \phi_0 - \Sigma^r,
\end{equation}
where $\Sigma^r$ is the rearrangement term arising due to the density-dependent couplings and is necessary to ensure thermodynamical consistency. The form is
\begin{align}\label{ret}
    \Sigma^r ={}& \sum_b \Bigg[ \frac{\partial g_{\omega b}}{\partial n_b} \omega_0 n_b + \frac{\partial g_{\rho b}}{\partial n_b} \rho_{03} I_{3b}  n_b+ \frac{\partial g_{\phi b}}{\partial n_b} \phi_0 n_b - \frac{\partial g_{\sigma b}}{\partial n_b} \sigma_0 n_b^s\Bigg],
\end{align}
The $\mu_b$ are determined by the chemical equilibrium condition
\begin{equation}
    \mu_b=\mu_n-q_b\mu_e, \label{beta}
\end{equation}
in terms of the chemical potential of the neutron and the electron, with $\mu_\mu=\mu_e$. The particle populations of each individual species are determined by Eq.~\eqref{beta} together with the charge neutrality condition of $\sum_i n_iq_i=0$, where $q_i$ is the charge of the baryon or lepton $i$.

\smallskip

The expression for pressure can be written in terms of the energy density and the chemical potential as
\begin{equation}
    P =\sum_i \mu_i n_i - \epsilon + n_B \Sigma^r,
\end{equation}
which receives a correction from the rearrangement term to guarantee thermodynamic consistency and energy-momentum conservation~\cite{Typel1999, PhysRevC.52.3043}.

\subsubsection{Deconfined quark matter}

We utilize the density-dependent quark mass (DDQM) model to represent quark matter. This straightforward and adaptable framework is ideal for exploring the deconfinement phase transition in hybrid stars, as referenced in \cite{backes2021effects}. The DDQM model emulates QCD quark confinement using quark masses that depend on density, defined by
\begin{equation}
    m_q = m_{q0} + \frac{D}{n_B^{1/3}} + Cn_B^{1/3} = m_{q0} + m_Q,
    \label{masses}
\end{equation}
where ${m_{q0}}$ (${q = u, d, s}$) corresponds to the current mass of the $q$th quark, ${n_B}$ is the baryon number density, and ${m_Q}$ corresponds to the density-dependent term that encompasses the interaction between quarks. The model parameters $D$ and $C$ dictate linear confinement and the leading-order perturbative interactions, respectively \cite{Xia2014}.

\smallskip

Careful treatment is needed to maintain thermodynamic consistency when introducing density dependence for state variables such as density, temperature, or magnetic field, similar to the method in Eq. \eqref{ret} for the DD-RMF model. We use the formalism from \cite{Xia2014} to ensure thermodynamic consistency in DDQM. At zero temperature, the basic energy density differential relationship is described as follows:
\begin{equation}
    \text{d}\varepsilon = \sum_q \mu_q\text{d}n_q,
    \label{diff-fundamental-eq}
\end{equation}
where $\varepsilon$ represents the matter contribution to the energy density of the system, ${\mu_q}$ are ${n_q}$ are the particle chemical potentials and particle densities, respectively. 

\smallskip

Expressing this model in terms of effective chemical potentials, we write the energy density for a free system as
\begin{equation}
    \varepsilon = \Omega_0 (\{\mu_q^*\},\{m_q\}) + \sum_q \mu_q^* n_q,
    \label{free-system-fundamental-eq}
\end{equation}
 using the density-dependent quark masses ${m_q(n_B)}$  and effective chemical potentials ${\mu_q^*}$.  The ${\Omega_0}$ is the thermodynamic potential of a free system. 
 Differentiating this form to yield
\begin{equation}
    \text{d}\varepsilon = \text{d}\Omega_0 + \sum_q \mu_q^* \text{d} n_q + \sum_q n_q \text{d}\mu_q^*.
    \label{initial-diff-free-system}
\end{equation}
We can then write d${\Omega_0}$ as
\begin{equation}
    \text{d} \Omega_0 = \sum_q \frac{\partial \Omega_0}{\partial \mu_q^*} \text{d}\mu_q^* + \sum_q \frac{\partial \Omega_0}{\partial m_q}\text{d}m_q,
\end{equation}
with
\begin{equation}
    \text{d}m_q = \sum_j \frac{\partial m_q}{\partial n_j} \text{d}n_j,
\end{equation}
where, in order to ensure thermodynamic consistency, the densities are connected to the effective chemical potentials by the relation
\begin{equation}
    n_q = -\frac{\partial \Omega_0}{\partial \mu_q^*}.
\end{equation}
Eq. \eqref{initial-diff-free-system} can thus be rewritten as
\begin{equation}
    \text{d}\varepsilon = \sum_q \left(\mu_q^* + \sum_j \frac{\partial \Omega_0}{\partial m_j} \frac{\partial m_j}{\partial n_q}\right) \text{d}n_q,
    \label{diff-free-system}
\end{equation}
providing a relation between the real and effective chemical potentials,
\begin{equation}
    \mu_q = \mu_q^* + \sum_j \frac{\partial \Omega_0}{\partial m_j} \frac{\partial m_j}{\partial n_q}.
\end{equation}
The pressure ${P}$ for the system can be obtained rom the fundamental relation $P = -\varepsilon + \sum_q \mu_q n_q$, as:
\begin{equation}
      P ={}-\Omega_0 + \sum_{q,j} \frac{\partial \Omega_0}{\partial m_j}n_q\frac{\partial m_j}{\partial n_q},
    \label{pressure-quarks}
\end{equation}
yielding a thermodynamically consistent EoS for quark matter.

\subsubsection{Phase transition}

The characteristics of the transition are influenced by the quark and hadron EoS models utilized. This study assumes the hadron-quark deconfinement transition to be a first-order phase transition, as effective models in the QCD phase diagram's high-density sector suggest. A phase transition can manifest as either a Maxwell \cite{PhysRevD.88.063001} or mixed (Gibbs) \cite{Glendenning:1992vb} phase transition, determined by the hadron-quark phase surface tension. In Maxwell construction, the phases remain distinct, ensuring local charge conservation, whereas, in Gibbs construction, quarks and hadrons coexist over a range of baryonic densities with global charge conservation. The thermodynamic process involves aligning the EoS of both phases and pinpointing the coexistence point. The method chosen for mixed-phase construction affects the stellar properties of the associated EoS \cite{Rather:2020lsg, Rather:2021yxo}.

\smallskip

For this work, we utilize the Maxwell construction to develop a hybrid EoS featuring a first-order phase transition at specific critical values of baryonic chemical potential and pressure. As per Gibbs' criteria, the transition takes place at the point where
\begin{align}
     P^{(i)}={}&P^{(f)}=P_0,\\
  \mu^{(i)}(P_0)={}&\mu^{(f)}(P_0)=\mu_0,
\label{eq:gibbscon}
\end{align}
sets the transition between the initial (${i}$) and final (${f}$) homogeneous phases with
\begin{equation}
 \mu^{ (i,f)}=\frac{\varepsilon ^{(i,f)}+P^{(i,f)}}{n_B^{(i,f)}},
\end{equation}
where ${\varepsilon ^{(i,f)}}$, ${P^{(i,f)}}$ and ${n_B^{(i,f)}}$ are the total energy density, pressure, and baryon number density, obtained from the EoS of each phase.
The conditions above the values of $P_0$ and $\mu_0$ are to be determined from the equations of state of both hadronic and deconfined quark phases. 
The transition point location, for a given baryonic composition in the hadronic phase, will be notably influenced by the choice of the free parameters for the DDQM model \cite{backes2021effects}.
In this study, we used a particular set of ($C, D^{1/2}$) for pure nucleonic EoS and another set for hypersonic EoS. The choice of these parameters is explained in \cite{Rather:2024hmo}.

\section{Numerical results and discussion}
\label{results}


\begin{figure}[h]
\centering
  \includegraphics[width=0.77\textwidth]{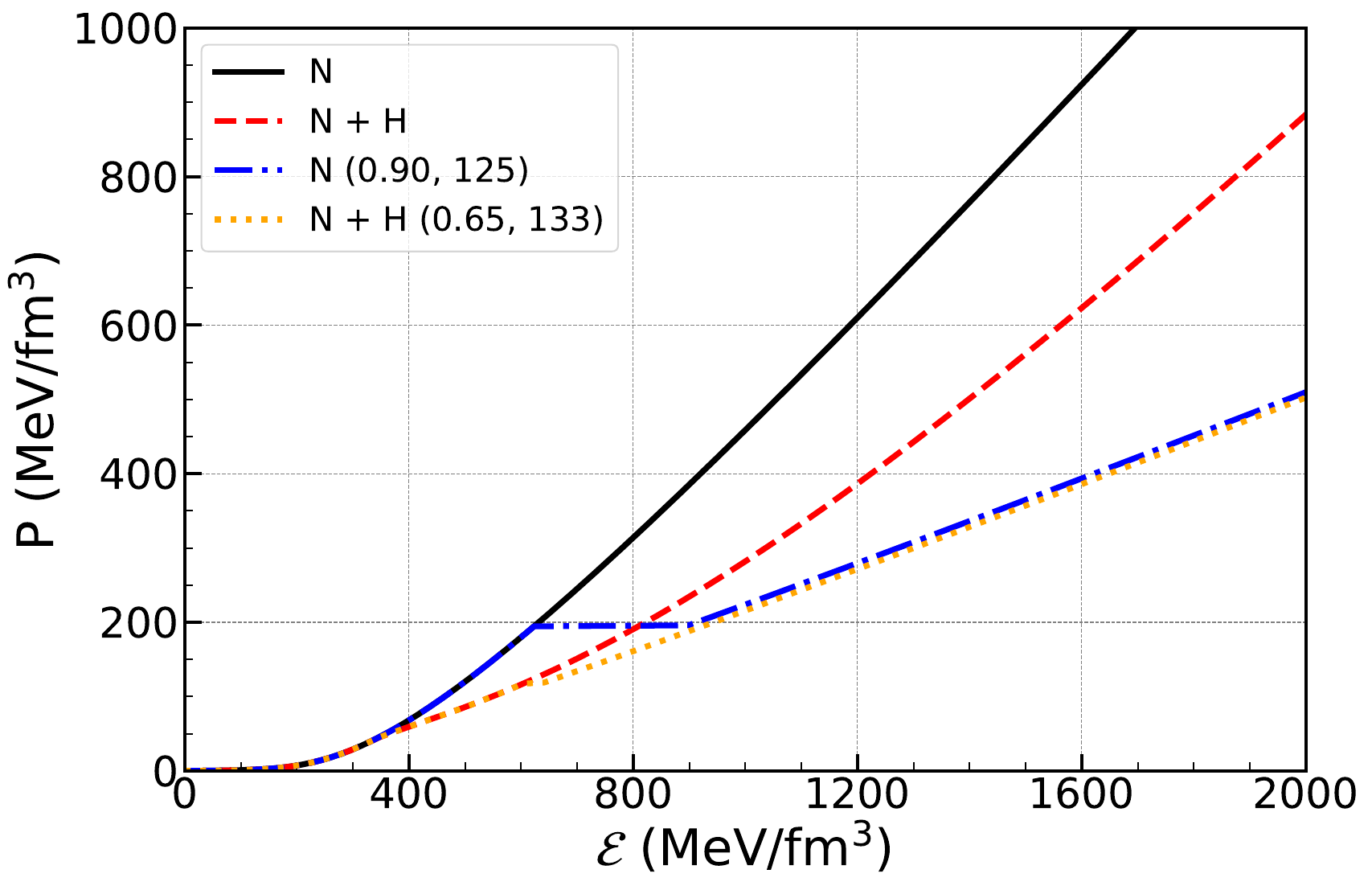}
			 			\caption{Energy density and pressure variation for the given DD-ME2 parameter set without and with a phase transition to the quark matter at different quark model parameters ($C, D^{1/2}$). The solid (dashed) line represents the pure nucleonic matter (nucleonic with hyperons) without a phase transition, while as dash-dotted (dotted) line represents the EoS for the nucleonic matter (nucleonic with hyperons) with a phase transition.}
		\label{figeos}	 	
\end{figure}

Figure~\ref{figeos} illustrates how pressure varies with energy density (EoS) for a neutron star under beta-equilibrium and charge-neutral conditions. The solid line corresponds to the EoS with nucleons only (N) while as the dashed line corresponds to the EoS with nucleons and hyperons (N + H). The dash-dotted and dotted lines correspond to the hybrid EoS with nucleons only, N (0.90, 125), and nucleons with hyperons, N + H (0.65, 133), where the numbers in the brackets correspond to the specific choice of quark model parameters ($C, D^{1/2}$). The EoS for the pure nucleonic EoS is very stiff implying that the rise in the pressure for a given energy density is very large. For the EoS with hyperons, N + H, we see that the low density part remains unchanged. The point where the EoS starts to deviate from the pure nucleonic one, around 380~MeV/fm$^3$, marks the appearance of hyperons that softens the EoS.

\smallskip

Regarding the hybrid EoSs, we see that the hybrid nucleonic EoS, N (0.90, 125), has a sharp transition where the pressure is constant, shifts to the quark matter EoS with a mixed phase region from around 600~MeV/fm$^3$ to 900~MeV/fm$^3$, a jump of 300~MeV/fm$^3$. The presence of hyperons causes a shift in the coexistence point towards lower density, marking a mixed-phase region of around 80~MeV/fm$^3$. Thus, for hybrid N EoS, the phase transition takes place at a very high density compared to hybrid N + H EoS. For the hybrid N + H EoS, the hadron-quark phase transition region is small and occurs at low density compared to the others. This implies a large quark phase present in comparison to the other hybrid EoSs. Post-phase transition, the EoS at higher densities is much more uniform compared to its hadronic counterpart. For instance, the parameter set (${C, D^{1/2}}$) = (0.90, 125 MeV) results in only a slightly stiffer EoS than (${C, D^{1/2}}$) = (0.65, 133 MeV). However, the position of the coexistence point plays the most crucial role when constructing the hybrid EoS. 

\smallskip

For a full unified EoS with crust part, we use the Baym-Pethick-Sutherland (BPS) EoS \cite{Baym:1971pw} for the outer crust, while the inner crust EoS is generated using the DD-ME2 parameter set in the Thomas-Fermi approximation \cite{PhysRevC.79.035804, PhysRevC.94.015808, rather2020effect}.

\begin{figure}[h]
\centering
\includegraphics[width=0.77\textwidth]{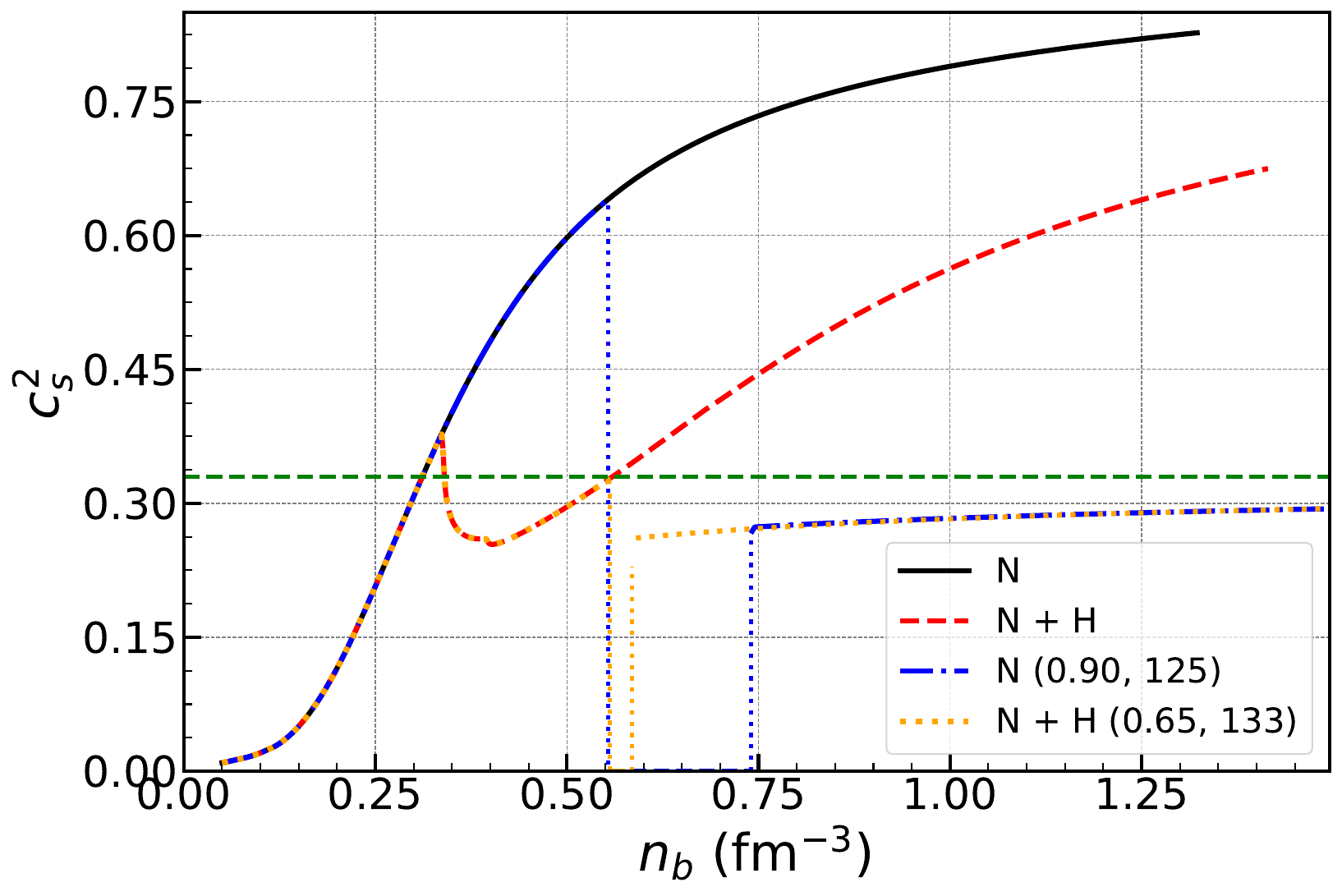}
			 			\caption{Speed of sound squared as a function of number density for the different hadronic composition EoS without (N, N + H) and with phase transition ( N (0.90, 125),  N + H (0.65, 133)) to the quark matter. The dotted lines in the right plot correspond to the mixed-phase region where $c_s^{2}$ drops to zero. The green dashed line in both plots represents the conformal limit $c_s^{2}$ = 1/3.}
		\label{figcs}	 	
     \end{figure}
  
Figure~\ref{figcs} shows the speed of sound squared as a function of number density for different compositions of the matter studied in this work, without and with a phase transition. 
From the thermodynamic stability, we need to ensure that $c_s^{2}$ $>$ 0 and from the causality, we have an absolute bound, $c_s^{2}$ $\leq$ 1. For very high densities, perturbative QCD findings anticipate an upper limit of $c_s^{2}$ = 1/3 \cite{PhysRevLett.114.031103}. The two solar mass requirements, according to several studies \cite{PhysRevLett.114.031103, PhysRevC.95.045801, Tews_2018}, necessitates a speed of sound squared that exceeds the conformal limit ($c_s^{2}$ = 1/3), revealing that the matter inside of NS is a highly interacting system.
In Figure~\ref{figcs}, the $c_s^{2}$ for pure nucleonic matter is significantly high, around 0.75, for the maximum mass configuration. In the appearance of different particles, such as hyperons, we can see the kinks corresponding to the onset of a new particle species, resulting in noticeable changes in the speed of sound squared. The value decreases to 0.54 for nucleonic EoS with hyperons. 

\smallskip

\begin{figure*}[t]
		\begin{minipage}[t]{0.49\textwidth}		 		
  \includegraphics[width=\textwidth]{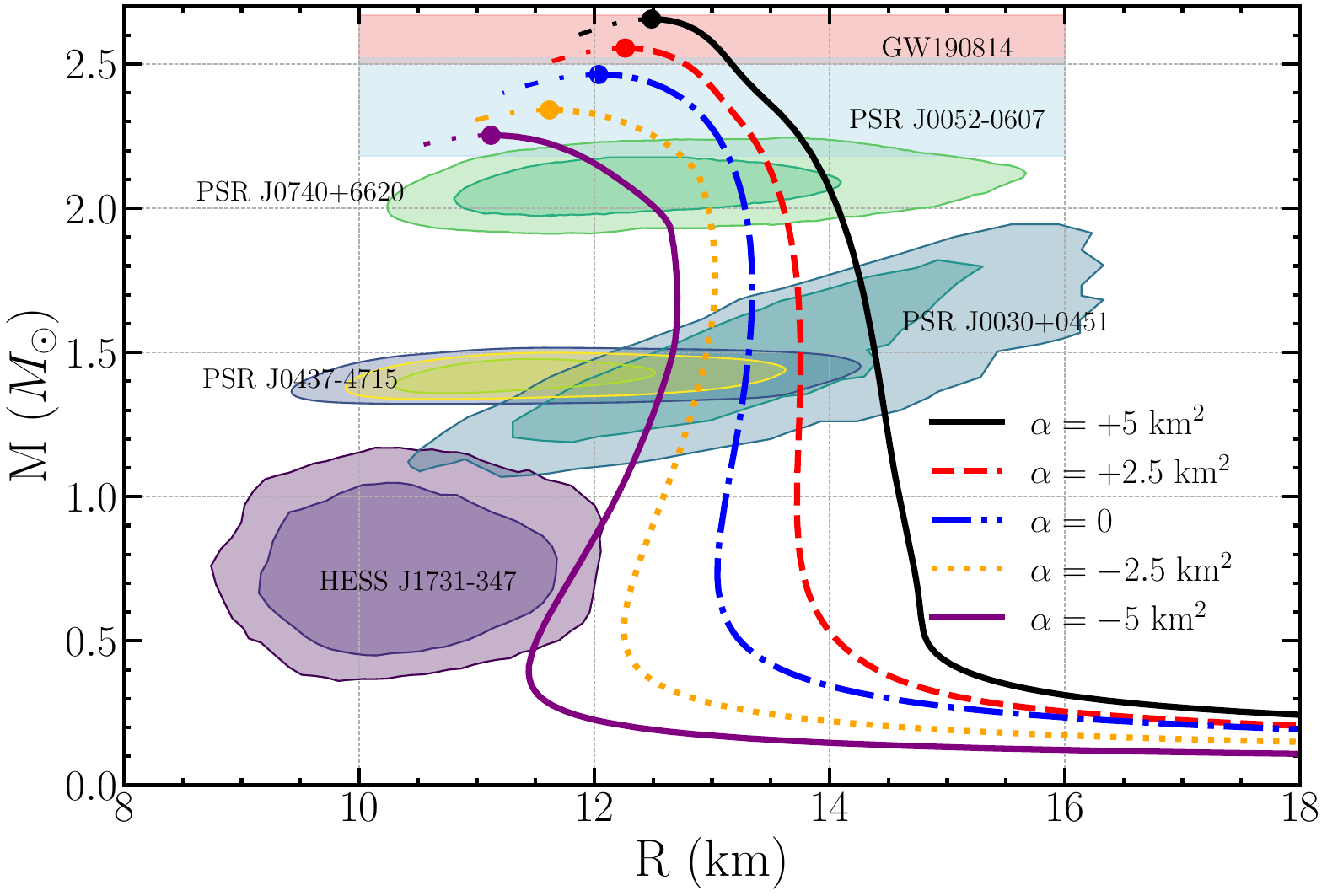}
			 	\end{minipage}
		 		\begin{minipage}[t]{0.49\textwidth}
			 		\includegraphics[width=\textwidth]{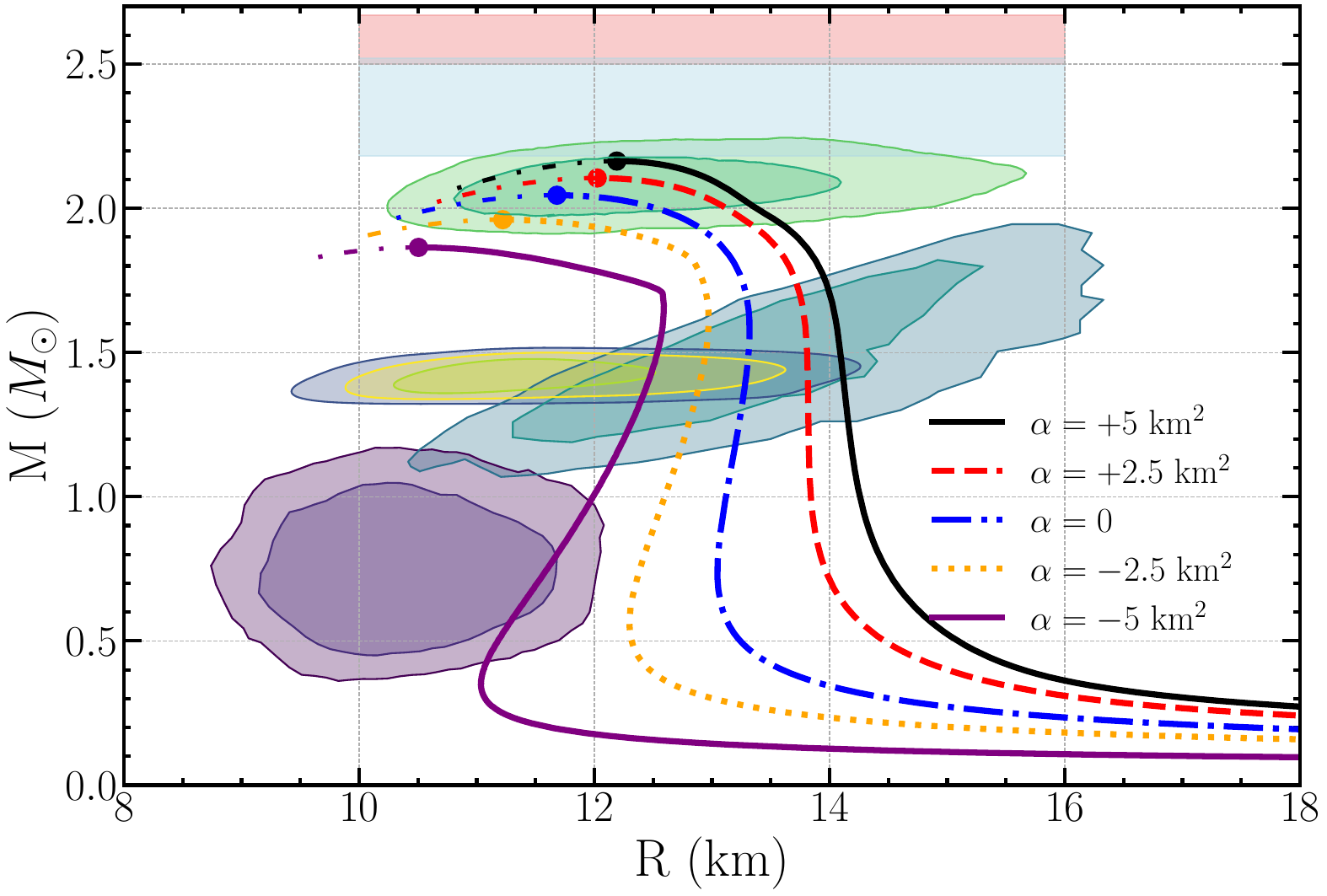}
			 	\end{minipage}
			 			\caption{Left: Mass-Radius relation for the nucleonic matter (left) and nucleons with hyperons (right) at different values of $\alpha$. The various shaded areas are credibility regions from the mass and radius inferred from the analysis of GW190814, PSR J0952-0607, PSR J0740+6620, PSR J0030+0451, PSR J0437-4715, and HESS J1731-347  as discussed in the text.} 
		\label{figmr1}	 	
\end{figure*}

    With the phase transition to the quark matter, $c_s^{2}$ drops to zero in the mixed-phase region because the pressure remains constant and then increases again in the pure quark phase. For hybrid EoS with nucleons only, we see that the speed of sound squared increases to higher values when it is still in the pure hadronic phase. Within the mixed-phase region, which extends from number density $n_b$ = 0.56 fm$^{-3}$ to $n_b$ = 0.74 fm$^{-3}$, the speed of sound squared remains zero. After the phase transition, it drops to a value of 0.27 at the maximum mass configuration, which lies in the pure quark phase. For the hybrid EoS with nucleons and hyperons, the mixed-phase region is very small, $n_b$ = 0.57-59 fm$^{-3}$, and the maximum speed of sound squared is 0.25. At high energy densities, all speed of sound value stays well below the conformal limit, unlike previous observations, due to the expected approach of a deconfined EoS towards the conformal limit from below.

 It is important to acknowledge a key simplification in our approach. The results presented here are based on the specific DDME2 parameterization for the hadronic EoS. As indicated in Table 2, the fundamental properties of nuclear matter, such as the symmetry energy slope $L_0$ and incompressibility $K_0$, are constrained by experiment only within a certain range of uncertainty. A different choice of parameters within these accepted ranges would yield a slightly different EoS, thus shifting the baseline ($\alpha$ = 0) mass-radius relations. 
We emphasize that our primary conclusions regarding the qualitative impact of the 4DEGB coupling constant $\alpha$ are robust. The modifications to the stellar structure equations introduced by $\alpha$ are driven by the gravitational theory and are independent of the specific EoS employed. Therefore, while the precise quantitative values for mass and radius are EoS-dependent, the general trend of positive $\alpha$ allowing for more massive and larger stars, and negative $\alpha$ leading to more compact configurations, is a fundamental feature of the 4DEGB gravity model.
    
Figure~\ref{figmr1} illustrates the mass-radius relationship based on solutions of the modified TOV equations for N (left) and N + H (right) EoSs at different values of the GB constant $\alpha$. The solid dot represents the last stable point reached in the center of the maximum-mass solution of the TOV equation. The dash-dotted line after the solid dot corresponds to the unstable part. In the left plot, for pure nucleonic matter, the maximum mass reaches 2.46\,$M_{\odot}$ with a radius of 12.04 km for $\alpha$ = 0, which resembles the MR relation obtained by solving the TOV equations for the equilibrium structure of an NS \cite{Rather:2023dom}. With the negative values of $\alpha$ (km$^2$), the MR relation shifts to a low radius and hence low maximum mass. For $\alpha$ = -2.5 and -5.0 km$^2$, the maximum mass decreases to 2.34 and 2.25\,$M_{\odot}$, respectively. The radius at 1.4\,$M_{\odot}$ decreases to 12.90 and 12.60 km, respectively, for $\alpha$ = -2.5 and 5.0 km$^2$.  While the standard $\alpha$ = 0 satisfies the astrophysical constraints from several measurements, the MR relation for $\alpha$ = -5.0 km$^2$ satisfies the low mass HESS J1731-347 constraint, thereby explaining its nature to be a hadronic star.
For positive values of GB constant, the maximum mass increases to a value of 2.55 and 2.65\,$M_{\odot}$ for $\alpha$ = +2.5 and +5.0 km$^2$, respectively. These values of the maximum mass satisfies the GW190814 constraint \cite{LIGOScientific:2020zkf} which lies in the so-called mass gap region. This suggests that the nature of the secondary component of GW190814 could be a supermassive NS. 

\smallskip
 
For the EoS with nucleons and hyperons, the maximum mass for $\alpha$ = 0 is 2.04\,$M_{\odot}$ with a radius of 13.28 km at 1.4\,$M_{\odot}$. For $\alpha$ = -2.5 and -5.0 km$^2$, the maximum mass decreases to 1.96 and 1.86\,$M_{\odot}$, respectively. These values lie well below the 2.0\,$M_{\odot}$ limit which is a requirement to describe the NSs. So in the case of N + H EoS, the negative values can be ignored in terms of satisfying the astrophysical constraints. For positive values, the maximum mass increases to a value of 2.10 and 2.16\,$M_{\odot}$ for $\alpha$ = +2.5 and +5.0 km$^2$, respectively. 

\begin{figure*}[h]
		\begin{minipage}[t]{0.49\textwidth}		 		
  \includegraphics[width=\textwidth]{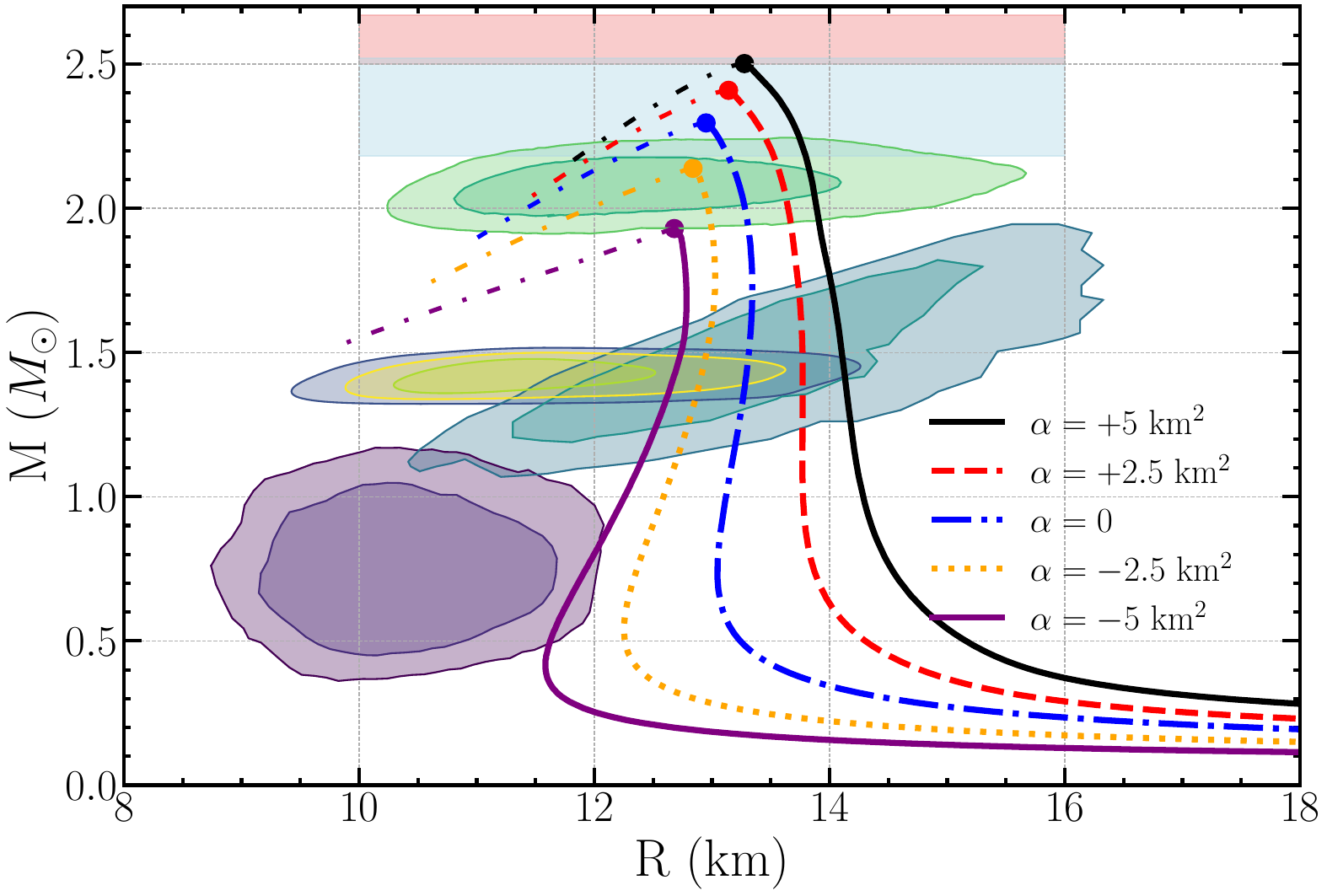}
			 	\end{minipage}
		 		\begin{minipage}[t]{0.49\textwidth}
			 		\includegraphics[width=\textwidth]{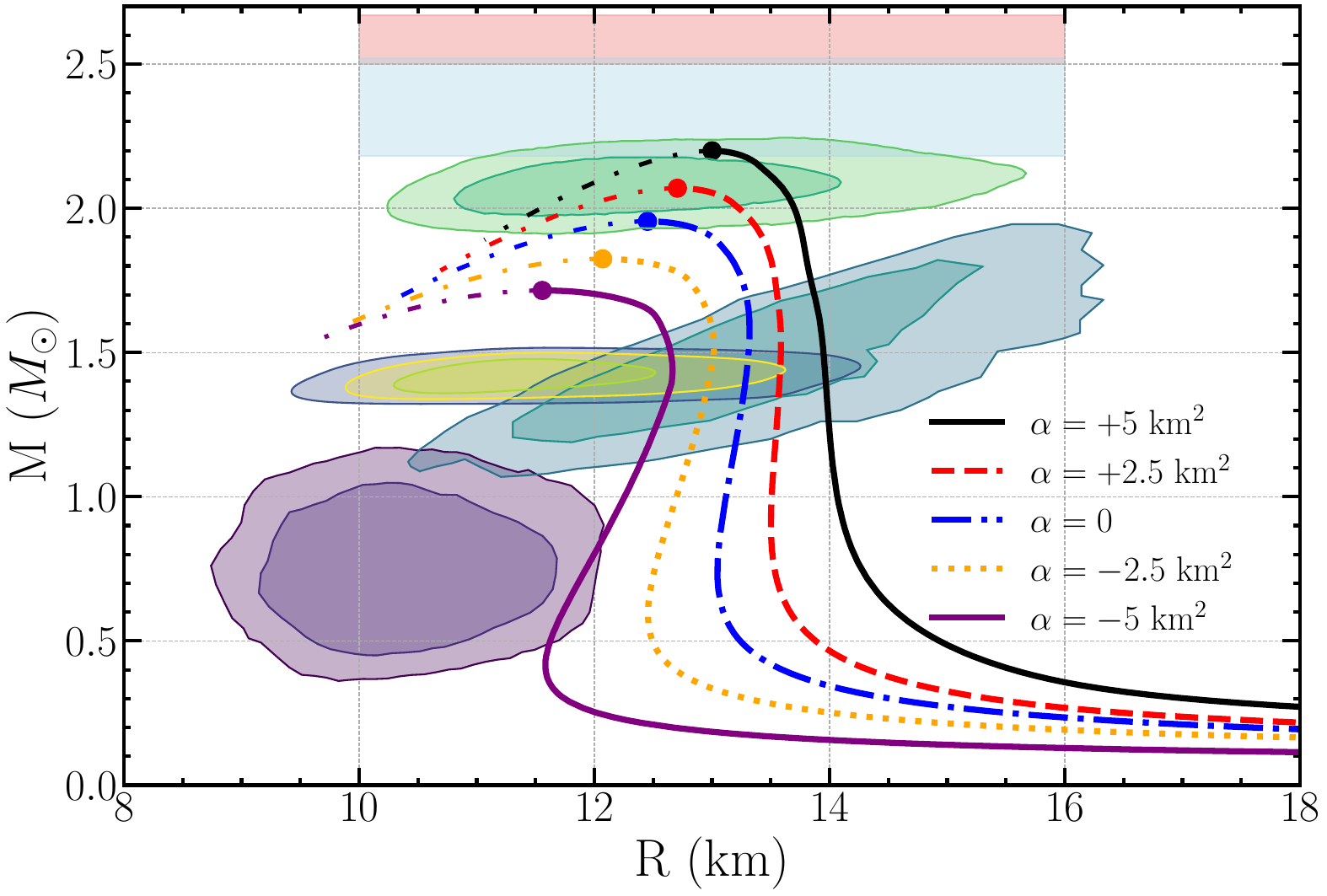}
			 	\end{minipage}
			 			 \caption{ Same as Figure \ref{figmr1}, but with a phase transition to the quark matter at different quark model parameters ($C$, $D^{1/2}$). }
		\label{figmr2}	 	
\end{figure*}

Figure \ref{figmr2} shows the mass-radius relation at different GB constant $\alpha$ for nucleonic (left) and hyperonic (right) EoS with a phase transition to the quark matter at different quark model parameters ($C$, $D^{1/2}$). For the hybrid EoS with nucleons only at $\alpha$ = 0, the maximum mass is 2.29\,$M_{\odot}$ with a radius of 13 km resembling the MR relation obtained by solving the TOV equations for the equilibrium structure of an NS \cite{Rather:2024hmo}. With negative values of $\alpha$, the maximum mass decreases to around 1.93\,$M_{\odot}$, lying below the 2.0\,$M_{\odot}$ limit. The radius at 1.4\,$M_{\odot}$ decreases to a value of 12.66 km for $\alpha$ = -5.0 km$^2$. With positive values, a maximum mass of 2.50\,$M_{\odot}$ and a radius of 14.13 km is obtained at 1.4\,$M_{\odot}$, satisfying all the necessary astrophysical constraints.

\smallskip

For the hybrid EoS with nucleons and hyperons, the maximum mass is 1.95\,$M_{\odot}$ with a radius of 12.54 km. Despite selecting quark parameters for a stiff EoS, including hyperons and a phase transition to quark matter leads to an EoS that softens enough to limit the star's maximum mass to slightly under 2\,$M_{\odot}$.  This is because of the initial low value of the maximum mass for the EoS without a phase transition. The maximum mass value decreases to 1.71\,$M_{\odot}$ for $\alpha$ = -5.0 km$^2$, satisfying the HESS J1731-347 constraint, and 2.20\,$M_{\odot}$ for $\alpha$ = +5.0 km$^2$, satisfying the necessary 2\,$M_{\odot}$ and other NICER measurements. In both the hybrid EoSs, the phase transition occurs at a very high density, allowing for a very small amount of quark matter in the core.
 
 The choice of the range $\alpha$ $\in$ [-5,+5] km$^2$ is further motivated by the astrophysical constraints themselves when considering all four EoS cases. For the hadronic EoSs (Fig. 3), the pure nucleonic case can satisfy the 2\,$M_{\odot}$ constraint for all values of $\alpha$ in this range. However, the inclusion of hyperons or a phase transition significantly softens the EoS, making negative values of $\alpha$ problematic. For the N+H EoS, $\alpha$= -5.0 km$^2$ already fails the 2\,$M_{\odot}$ constraint with a maximum mass of only 1.86\,$M_{\odot}$. Similarly, for the hybrid EoSs (Fig. 4), values of $\alpha$= -5.0 km$^2$ result in maximum masses of 1.93\,$M_{\odot}$ (nucleonic) and 1.71\,$M_{\odot}$ (hyperonic), both falling short of the required 2\,$M_{\odot}$. Therefore, a more negative value, such as $\alpha$= -6 km$^2$, would be ruled out for three of the four EoSs considered. Conversely, while positive values of $\alpha$ help satisfy the mass constraint, they also increase the stellar radius. For instance, the hybrid nucleonic case with $\alpha$= +5.0 km$^2$ predicts a radius of 14.13 km at 1.4\,$M_{\odot}$. A larger positive value, such as $\alpha$= +6 km$^2$, would push the radius even further, potentially violating the upper bounds from NICER measurements for pulsars like PSR J0740+6620 and PSR J0030+0451. Thus, the chosen range represents a phenomenologically interesting window where the interplay between modified gravity and the underlying EoS can be effectively studied against current observational bounds.
  
\smallskip

The above plots showed how the GB constant $\alpha$ affects the overall MR relation at certain values. In order to see a more general behavior, we calculated the MR relation for several values of $\alpha$ = 0, -1.0, -2.0, -2.5, -3.0, -4.0, -5.0 km$^2$, and the corresponding positive values. We calculated the maximum mass and the corresponding radius for all the values, and plotted them with some fit functions, to see a general behavior.

\smallskip

\begin{figure}[h]	 	
\centering
  \includegraphics[width=0.70\textwidth]{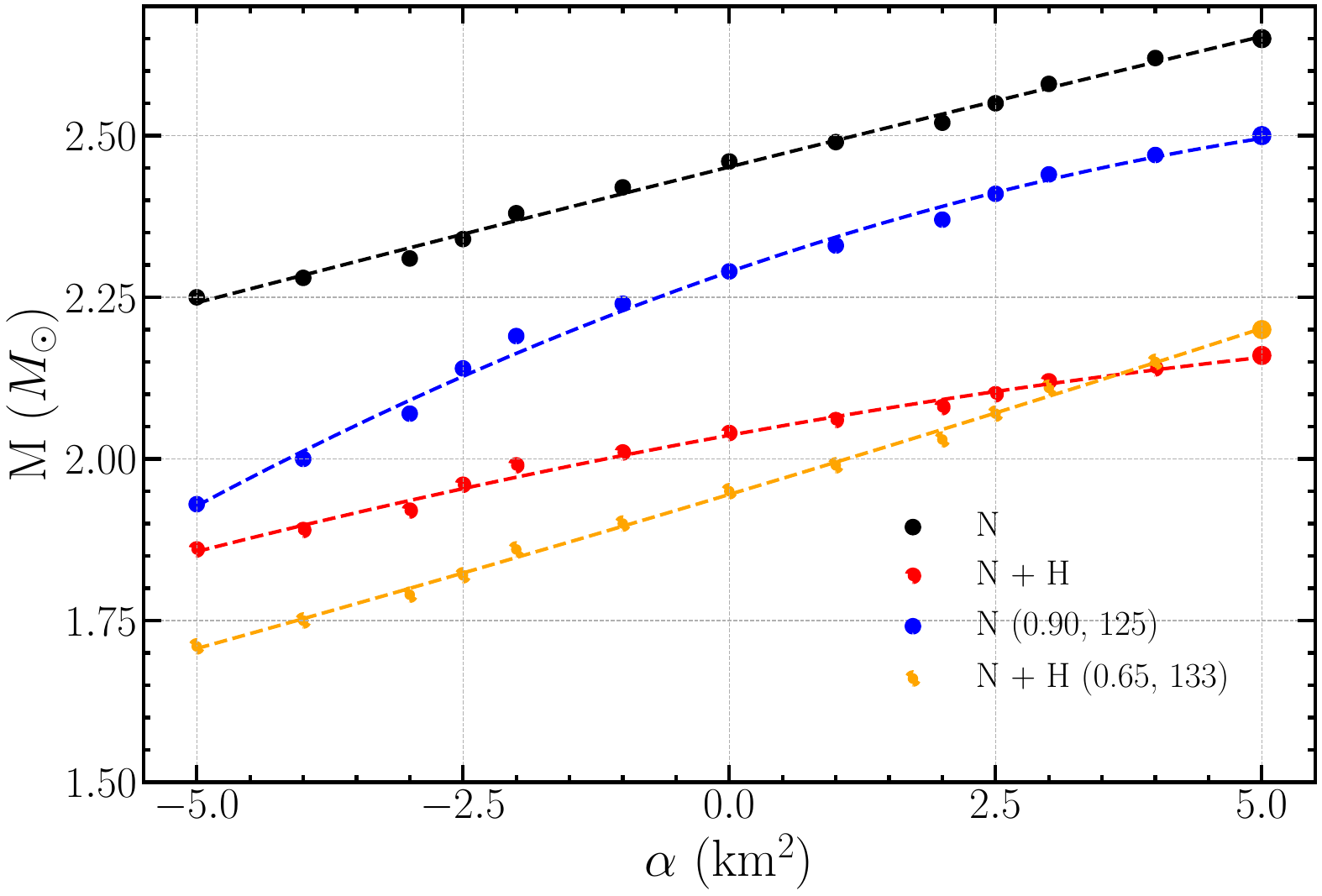}
			 			 \caption{ Variation of the maximum mass for different compositions of the EoS without and with phase transition at different values of the GB constant $\alpha$. }
		\label{figam}	 	
\end{figure}

Figure \ref{figam} displays the variation of the maximum mass for different compositions of the EoS without and with phase transition, at different values of the GB constant $\alpha$. We see that the maximum mass increases for positive values of $\alpha$ and vice-versa for negative values. The change in the maximum mass for nucleonic EoS without and with phase transition keeps increasing with the increase in the value of $\alpha$. For the hyperonic EoS at higher values, with and without a phase transition, the maximum mass changes slightly. At $\alpha$ = +4.0 and +5.0 km$^2$, the hybrid EoS results in more massive stars than those without a phase transition.

\smallskip

To fit the function between maximum mass and the constant $\alpha$, we use the following form
\begin{equation}\label{fit:am}
M = a. (k.x)^2 + b.(k.x) +c
\end{equation}

Here $k$ is the scaling factor. As we can see form the plot, all the EoS satisfy the fit functions pretty accurately. The values of the constants and scaling factors for different EoSs are shown in Table \ref{T4}.

\begin{table}[h]
\centering
\caption {Values of fitting coefficients for Eq. \ref{fit:am}.
\label{T4}}
\begin{tabular}{ c p{1.7cm}p{1.2cm}p{0.9cm}p{1.9cm} } 
\hline
 EoS & $a$ ($M_{\odot}$) & $b$ ($M_{\odot}$) & $c$ ($M_{\odot}$) & $k$ (km$^{-2}$) \\
 \hline
N & 1.08 $\times$ 10$^{2}$ & 33.74 & 2.45 & 1.22 $\times$ 10$^{-3}$  \\  
 N + H & 3.69 $\times$ 10$^{3}$ & 53.08 & 2.04 & 5.66 $\times$ 10$^{-4}$ \\
N (0.90, 125) & 2.67 $\times$ 10$^{2}$ & 16.67 & 2.29 & 3.4 $\times$ 10$^{-3}$ \\
  N + H (0.65, 133)  & 5.59 $\times$ 10$^{1}$ & 19.26 & 1.94 & 2.57 $\times$ 10$^{-3}$  \\
  \hline
\end{tabular}
\end{table}

The same analysis is performed for the maximum radius at different values of $\alpha$. For the radius, we fit the following function:
\begin{equation}\label{fit:aR}
R = a. (k.x)^2 + b.(k.x) +c
\end{equation}

\begin{figure}[h]	 
\centering
  \includegraphics[width=0.70\textwidth]{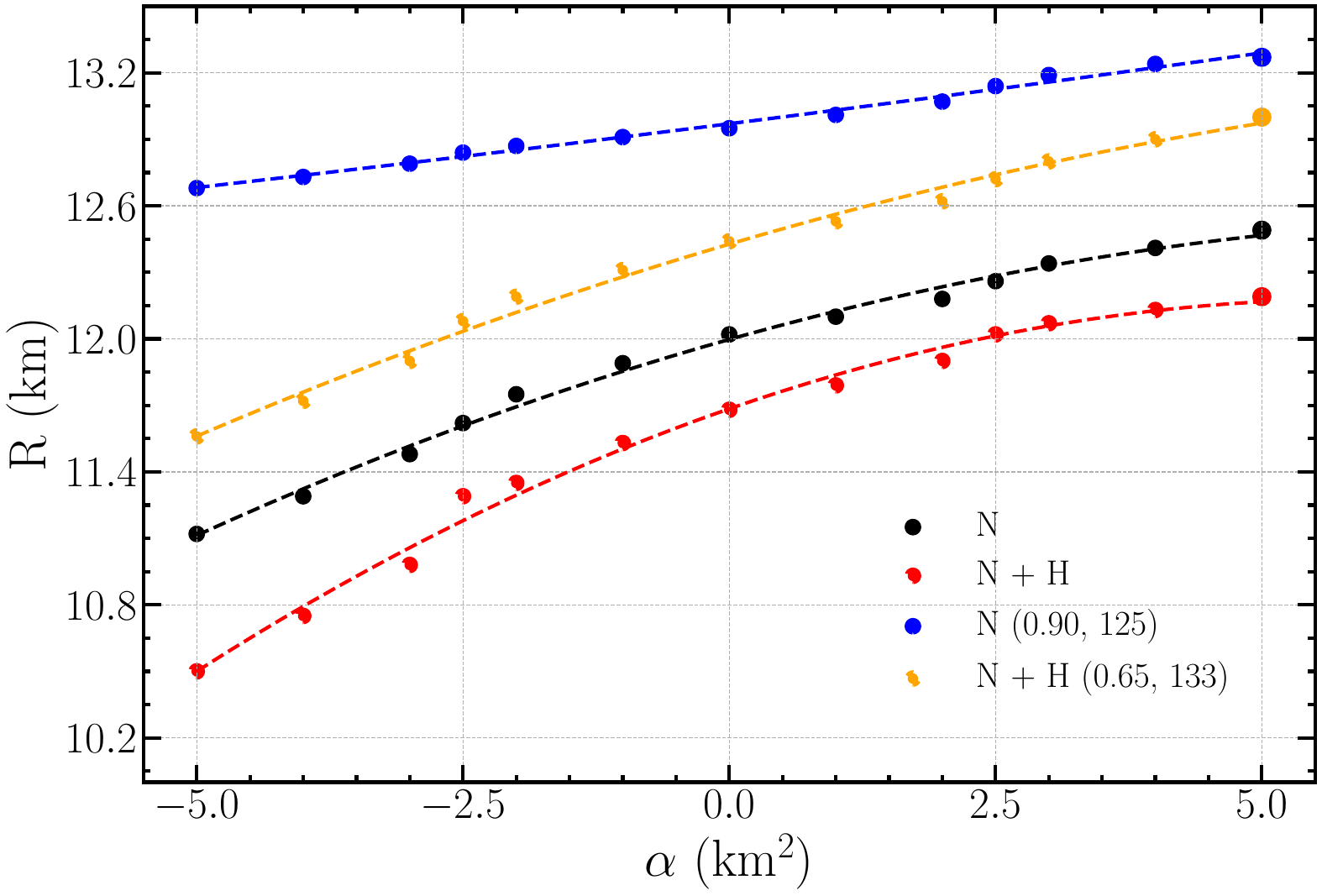}
			 			 \caption{ Same as Figure \ref{figam}, but for the maximum radius. }
		\label{figar}	 	
\end{figure}

Figure \ref{figar} shows the change in the maximum radius for different compositions of the EoS without and with phase transition, at different values of the GB constant $\alpha$. From the plot, we see that the initial radius for hybrid EoSs at $\alpha$ = 0 is higher than the EoS without a phase transition. For pure nucleonic EoS without a phase transition, the radius varies from 11.12  to 12.49 km for  $\alpha$ = -5.0 and +5.0 km$^2$, respectively. For the hybrid nucleonic EoS, this changes from 12.68 km to 13.27 km, respectively. So while the maximum radius for hybrid nucleonic EoSs is higher than the normal nucleonic EoS, the change in the radius is large for EoSs without a phase transition. Similar behavior is seen for the EoS with hyperons. The values of different constants and the scaling factor for different EoSs are shown in Table \ref{T5}.

\begin{table}[!ht]
\centering
\caption {Same as Table \ref{T4}, but for Eq. \ref{fit:aR}.
\label{T5}}
\begin{tabular}{ c p{1.7cm}p{1.7cm}p{0.9cm}p{1.9cm} } 
\hline
 EoS & $a$ (km$^{-1}$) & $b$ & $c$ (km) & $k$ (km$^{-1}$) \\
 \hline
N & 3.39 $\times$ 10$^{2}$ & 2.74 $\times$ 10$^{1}$ & 12.0 & 4.94 $\times$ 10$^{-3}$  \\  
 N + H & 5.58 $\times$ 10$^{2}$ & 3.33 $\times$ 10$^{1}$ & 11.68 & 5.01 $\times$ 10$^{-3}$ \\
N (0.90, 125) & 1.73 $\times$ 10$^{4}$ & 3.15 $\times$ 10$^{2}$ & 12.97 & 1.92 $\times$ 10$^{-4}$ \\
  N + H (0.65, 133)  & 1.49 $\times$ 10$^{4}$ & 2.15 $\times$ 10$^{2}$ & 12.43 & 6.56 $\times$ 10$^{-4}$  \\
  \hline
\end{tabular}
\end{table}

\subsection{Impact of Anisotropy}

In our initial analysis, we assumed the stellar matter to be a perfect isotropic fluid, where the radial pressure $P_r$ equals the tangential pressure $P_t$. However, physical conditions in the dense core of a NS, such as strong magnetic fields, superfluidity, or phase transitions, can lead to local pressure anisotropy \cite{Bowers:1974tgi}. To investigate this effect, we introduce an anisotropy measure $\Delta$ defined as the difference between the tangential and radial pressures:
\begin{equation}
\Delta(r) = P_t(r) - P_r(r).
\end{equation}
This modifies the hydrostatic equilibrium equation by introducing an additional term. We adopt the widely used model by Bowers and Liang \cite{Bowers:1974tgi}, where the anisotropy is related to the stellar compactness:
\begin{equation}
\Delta(r) = \kappa \frac{2m(r)}{r} P_r(r).
\end{equation}
Here, $\kappa$ is a dimensionless parameter that quantifies the magnitude and nature of the anisotropy. A positive value ($\kappa >$ 0) corresponds to a repulsive anisotropic force that makes the EoS effectively stiffer, while a negative value ($\kappa <$ 0) corresponds to an attractive force that softens the EoS. The modified TOV equation for the radial pressure gradient (Eq. \ref{dpdr}) then includes an additional term +2$\Delta/r$.

\begin{figure*}[htbp!]
\centering
		\begin{minipage}[t]{0.70\textwidth}		 		
  \includegraphics[width=\textwidth]{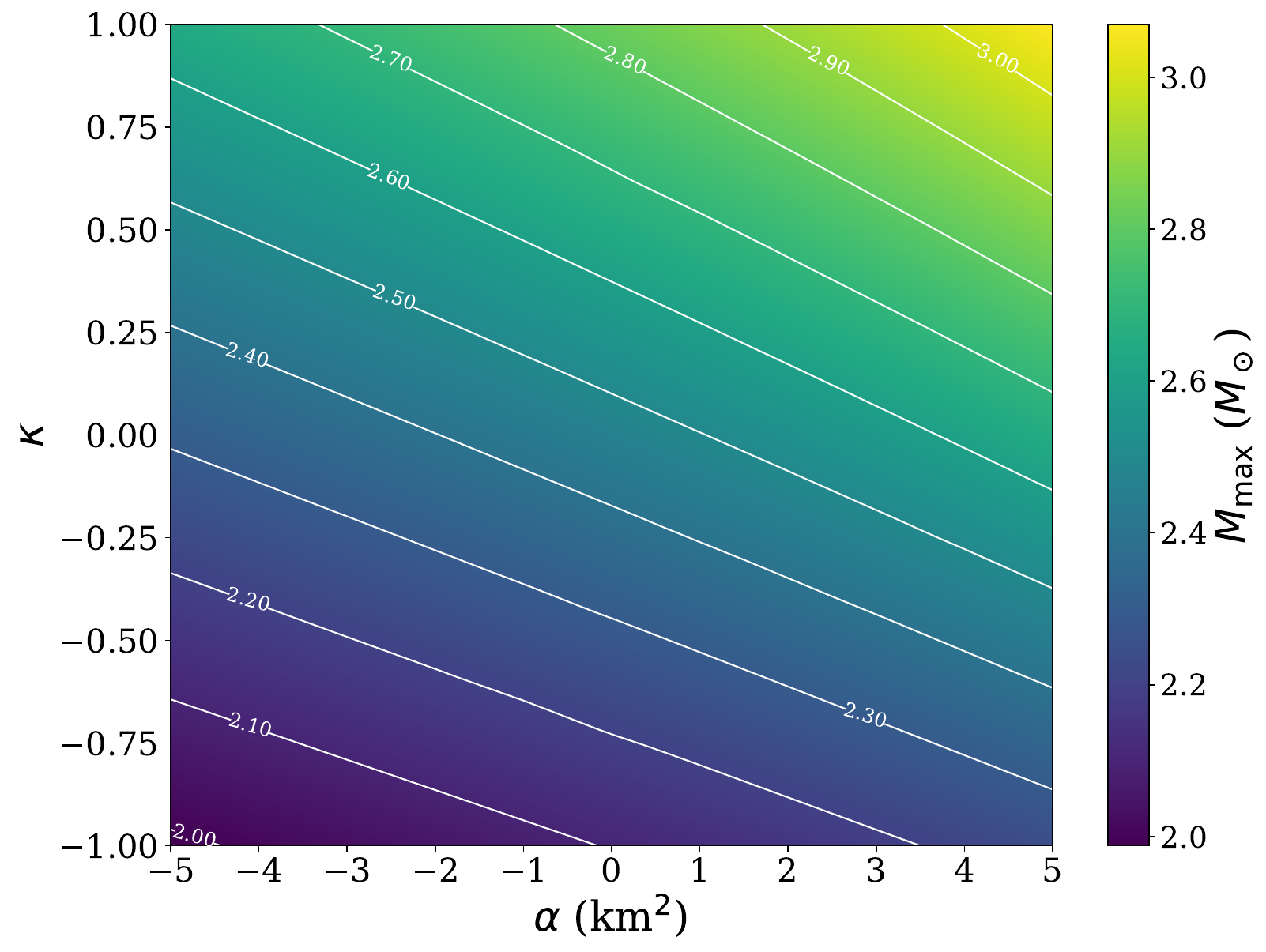}
			 	\end{minipage}
		 		\begin{minipage}[t]{0.70\textwidth}
			 		\includegraphics[width=\textwidth]{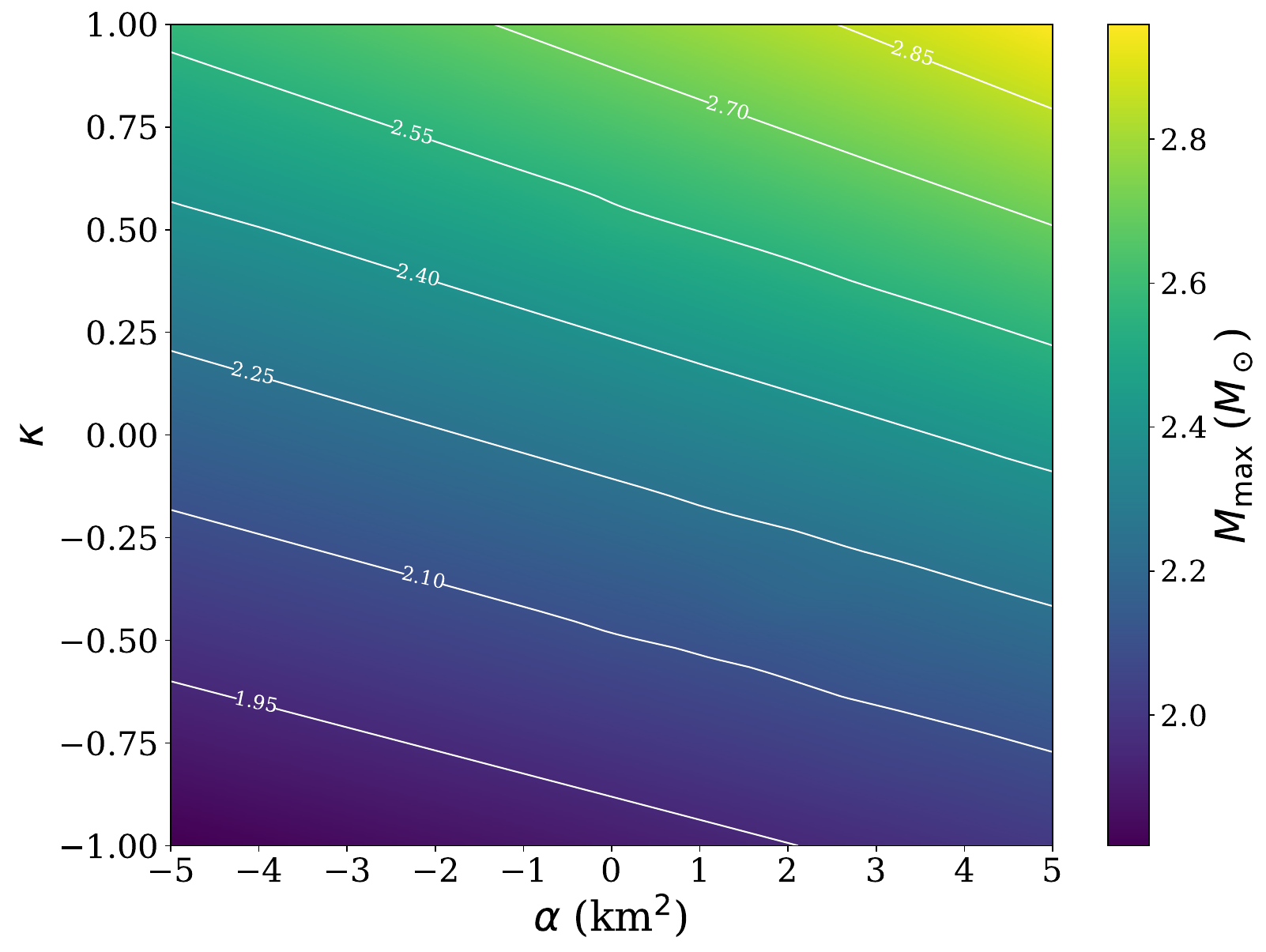}
			 	\end{minipage}
			 			 \caption{ Contour plots of the maximum stellar mass ($M_{max}$) in solar masses as a function of the EGB coupling constant $\alpha$ and the anisotropy parameter $\kappa$. The top panel shows the results for the purely nucleonic (N) EoS, while the bottom panel corresponds to the hybrid EoS with a phase transition to the quark matter, N (0.90, 125).}
		\label{figcontour}	 	
\end{figure*}

We have systematically studied the combined effects of the EGB coupling constant $\alpha$ and the anisotropy parameter $\kappa$ on the maximum mass of the star. The results are summarized in the contour plots shown in Fig.~\ref{figcontour} for the purely nucleonic EoS, N, and the hybrid EoS with a phase transition, N (0.90, 125).

Several key features are evident from both plots. The maximum mass consistently increases with more positive values of $\alpha$ and decreases for negative $\alpha$, in line with our findings for the isotropic case. More importantly, the anisotropy has a very strong effect: a positive (repulsive) $\kappa$ significantly raises the maximum mass, whereas a negative (attractive) $\kappa$ lowers it. The contour lines of constant mass illustrate a clear degeneracy between the modified gravity effect ($\alpha$) and the internal matter physics ($\kappa$). For instance, a reduction in the maximum mass caused by a negative $\alpha$ can be entirely compensated by introducing a sufficiently repulsive anisotropy ($\kappa >$ 0).

Comparing the two panels in Fig.~\ref{figcontour}, we observe that the overall mass scale is lower for the hybrid EoS. This is expected, as the phase transition softens the EoS, reducing the maximum mass of the star for any given ($\alpha$,$\kappa$) pair. However, the qualitative dependence on $\alpha$ and $\kappa$ remains remarkably similar, as shown by the nearly parallel slope of the contour lines in both plots. This suggests that the impact of anisotropy is a robust mechanism for modifying stellar structure, independent of the underlying stiffness of the EoS.

While these plots are for a purely nucleonic composition, the results can be directly extrapolated to the case with hyperons. The inclusion of hyperons further softens the EoS, which would shift the entire contour map to lower maximum mass values. In that scenario, a repulsive anisotropy ($\kappa >$ 0) could serve as a crucial physical mechanism to counteract the softening effect of hyperons, helping to reconcile theoretical models with the observational requirement of supporting massive, 2.0\,$M_\odot$ NSs.

\section{Summary and Conclusion}
\label{summary}

To summarize our work, in the present study we have investigated in detail the impact of hyperons on the structural properties of hadronic and hybrid stars within the framework of the regularized Einstein-Gauss-Bonnet theory of gravity in four-dimensional space-time. We briefly presented the action of the modified theory of gravity, the corresponding field equations, as well as the modified TOV equations for static, spherically symmetric stars. We discussed and described the formalism to obtain a realistic hadronic EoS compatible with astrophysical constraints, including hyperons within the relativistic mean field theory, which ensures that causality is never violated. Deconfined quark matter and phase transition were discussed as well. We have integrated the equations numerically, assuming both negative and positive values of the Gauss-Bonnet coupling constant within its allowed range, and we have displayed our main numerical results in several tables and figures.

\smallskip 

Our results show that the inclusion of hyperons softens the EoS, while the phase transition softens it even further. The speed of sound squared throughout the density range remains always positive and lower than unity, avoiding causality. As far as the mass-radius relationships are concerned, we observed that i) For a given stellar mass, a positive Gauss-Bonnet coupling constant $\alpha$ leads to more massive stars that can satisfy several astrophysical constraints which are otherwise difficult to fulfill with a standard EoS under general relativity ($\alpha = 0$). Negative values of $\alpha$, on the other hand, yield more compact stars. Although such configurations are consistent with certain low-mass and small-radius measurements, such as that of HESS J1731-347, they typically fail to satisfy the 2\,$M_{\odot}$ mass constraint and hence can be ruled out in the case of EoSs involving phase transitions.
ii) Depending on the specific stellar properties resulting from a given EoS in standard TOV gravity, the coupling constant $\alpha$ can be effectively constrained using the astrophysical measurements discussed in the text. For EoSs without phase transitions, small negative values of $\alpha$ remain well within current observational bounds, while all positive values are consistent with both the 2\,$M_{\odot}$ limit and the NICER radius constraints at 1.4\,$M_{\odot}$. However, in the case of EoSs featuring a strong first-order phase transition, negative values of $\alpha$—despite being compatible with observed radius constraints—fail to produce stars with masses above 2\,$M_{\odot}$, and can therefore be ruled out based on current mass measurements. This highlights the potential of gravitational theories beyond general relativity, such as Einstein-Gauss-Bonnet gravity, to be tested and constrained by precise astrophysical observations. 
iii) Fitting to the maximum mass and corresponding radius for several values of constant $\alpha$ provides a general behavior of change in these properties.

\smallskip

In addition to the modified gravity effects, we also explored the impact of local pressure anisotropy on the stellar structure. Our analysis shows that anisotropy provides a significant mechanism for altering the maximum mass, with a repulsive force ($\kappa >$ 0) increasing it and an attractive force ($\kappa <$ 0) decreasing it. We found a notable degeneracy between the effects of the EGB coupling constant $\alpha$ and the anisotropy parameter $\kappa$, where the influence of one can be counteracted by the other. This finding is particularly relevant for softer equations of state, as a repulsive anisotropy could help models with hyperons or strong phase transitions to satisfy the stringent two-solar-mass observational constraint.

\smallskip

 Regarding future work, this study lays the essential groundwork for a number of timely investigations. A key extension would be to compute the tidal deformability of compact stars within 4DEGB gravity. This involves deriving and solving the static perturbation equations for the metric and fluid variables to determine the second Love number $k_2$ and the corresponding dimensionless tidal deformability. Such a calculation would enable direct comparison with gravitational wave constraints, such as those from GW170817, providing a new avenue to constrain the coupling constant $\alpha$ using observables from the late-inspiral phase of binary mergers. This would also allow for an investigation into whether the quasi-universal relations between $\Lambda$, the moment of inertia, and compactness that hold in GR are modified by 4DEGB gravity.

A second, parallel line of inquiry involves incorporating stellar rotation, initially in the slow-rotation approximation to compute the moment of inertia and frame-dragging effects, and eventually via full numerical solutions for rapidly rotating stars. This would allow for a more direct connection to pulsar observations. Finally, the intriguing possibility of dark matter admixed neutron stars could be explored within this framework, investigating how the interplay between a new matter sector and modified gravity affects the stellar properties and potential observational signatures.

\section*{Acknowledgments}
I.A.R. acknowledges support from the Alexander von Humboldt Foundation.

\section*{ORCID}

\noindent Ishfaq Ahmad Rather - \url{https://orcid.org/0000-0001-5930-7179}

\noindent Grigoris Panotopoulos- \url{https://orcid.org/0000-0003-1449-9108}


\end{document}